# Understanding binary phase separation of Cu-C nanocrystalline-amorphous composites

*Jiajian Guan*,[†1] *Qin Jiang*,[†2] *Yue He*,[†1] *Xu Zhang*,[2,*] *Bin Liao*,[2] *Wei Gao*[1] and *Lizhao Qin*[3]

[1] Department of Chemical and Materials Engineering, Faculty of Engineering, the University of Auckland, PB 92019 Auckland, New Zealand

[2] Key Laboratory of Beam Technology and Materials Modification of Ministry of Education, College of Nuclear Science and Technology, Beijing Normal University, Beijing 100875, China

[3] School of Materials and Energy, Southwest University, Chongqing 400715, China

[*] Authors to whom correspondence should be addressed: zhangxu@bnu.edu.cn

[†] These authors contributed equally

**ABSTRACT**: The nanocrystalline-amorphous textures are commonly observed in the coatings synthesized by energetic deposition. This work reports a theoretical study towards binary Cu-C phase separation. By performing a MD simulation using the LAMMPS instead of classical PFK methodology, we theoretically explained how the initial pressure and Cu concentration fundamentally determines Cu nanocrystalline's final morphology and grain size, which gives a novel insight into binary phase separation and the evolution mechanism of nanocrystalline-amorphous structures during energetic deposition.

## I. INTRODUCTION

Binary phase separation is a very fundamental phenomenon in liquid-solid and solid-solid interfaces during fluid solidification from an elevated homogeneous temperature quenching to its critical solution temperature, particularly of metal-metal and metal-nonmetal types. For instance, binary metal composites like Cu-Ag [1], Cu-Ta [2], and Cu-Mo [3], and binary metal-nonmetal composites such as Cu-C [4], Ni-C [5], and Cr-C [6]. The increasing total free energy of the system due to small concentration fluctuations leads to phase separation by discontinuous nucleation and subsequent growth. This phenomenon is commonly observed not only in thermodynamic equilibrium states but also non-equilibrium structures like the metastable nanocrystalline-amorphous composites that fabricated by energetic deposition such as filtered cathodic vacuum arc (FCVA), high power impulse magnetron sputtering (HiPIMS), and direct current (DC) magnetron sputtering, in which long-range atomic diffusion has been rigidly confined due to the rapid heating and quenching processes at the atomic scale, impeding the final crystal growth during deposition [7]. These nanocrystalline-amorphous textures are considered as one of the significant factors contributing to the extraordinary mechanical properties of superhard composite coatings [8,9]. However, an atomic understanding of the phase separation from homogeneous blends to these nanocrystalline-amorphous coexisting structures remains unclear.

The shape, distribution, and solute concentration of the microstructures during binary phase separation can be mathematically predicted by phase field kinetic (PFK) model like classical Cahn-Hilliard (CH) theory as shown in Eq. (1), where parameter $M$, $F$, and $k$ represent the kinetic coefficient of diffusion (mobility), total free energy, and gradient energy coefficient respectively [10,11]. By using the PFK model, general spinodal decomposition [12-14] and its resulting structures such as lateral concentration modulations (LCMs) and vertical concentration modulations (VCMs) during film deposition can be theoretically reproduced [15-18]. However, unlike typical molecular dynamic (MD) simulation, the PFK model can't further give neither an atomic insight nor the environment details like temperature and pressure during phase separation. On the other hand, despite the spot-shaped nanocrystalline is commonly observed by energetic deposition, and its average size will decrease with increasing content of amorphous phase [19-21], little theoretical investigation has been focused on this phenomenon. Therefore, in order to unveil the key factors that fundamentally determine the morphology and size of crystalline during phase separation, we perform a typical MD simulation towards binary Cu-C nanocrystalline-amorphous composites, which may provide a more profound understanding than that from the PFK methodology.

$$\frac{\partial c}{\partial t} = M\nabla^2 \left(\frac{\partial F}{\partial c} - k\nabla^2 c\right) \quad (1)$$

## II. SIMULATION DETAILS

To reproduce the Cu-C phase separation at an atomic scale, general MD simulation was performed by using the large-scale atomic/molecular massively parallel simulator (LAMMPS) [22,23]. $10^4$ random Cu-C atoms with 0.5 g/cm$^3$ cell density were first constructed with the Materials Studio (MS) package (Version 8.0) as an estimated amorphous structure at the early stage of deposition. Secondly, the amorphous structure was quenched from 1000 K to 300 K under the NPT ensemble of the Nose-Hoover thermostat for 1 ns with a time step of 1 fs using the LAMMPS. The hybrid Cu-Cu, Cu-C, and C-C interactions were described by the embedded atom method (EAM) potential [24], Lennard-Jones (LJ) potential, and the adaptive intermolecular reactive empirical bond order (AIREBO) potential [25], respectively. By altering the pressure and atomic concentration gradient, we quantitatively investigated the evolution of Cu atomic diffusion, nucleation, and crystal growth. The mean-squared displacement (MSD) of Cu atoms within $\Delta t$ is given by Eq. (2), where $r(t)$ represents atomic position at time $t$ in the NPT ensemble, and the corresponding atomic diffusion coefficient is derived from Eq. (3). All the MD visualizations were implemented by using the OVITO Basic software [26].

$$MSD(\Delta t) = \langle [r(t+\Delta t) - r(t)]^2 \rangle \quad (2)$$

$$D = \frac{1}{6} \lim_{\Delta t \to \infty} \frac{dMSD}{d\Delta t} \quad (3)$$

## III. RESULTS AND DISCUSSION

Fig. 1 a-b present the snapshots of Cu-C atomic displacement trajectory with initial Cu:C=1:1 quenching from 1000K to 300K under 1GPa, indicating the hybrid Cu-C atoms will spontaneously form Cu-rich and C-rich phases during atomic diffusion. The process of atomic diffusion can be further revealed by recording the MSD of Cu and C atoms as shown in Fig. 1c. The results highlight three possible stages during atomic diffusion from 1000K to 300K (Fig. 1d). In the first stage (0-10ps), the diffusion coefficient of C is slightly larger than Cu with the value ranging from ~15 to ~24 $Å^2$/ps, and the increasing Cu content will boost the diffusion ability of both Cu and C in this stage. In the second stage (10ps-200ps), the Cu and C diffusion coefficients dramatically fall below ~0.11 $Å^2$/ps with C dropping faster than Cu, and the diffusion ability of C gradually increases with rising Cu concentration ratio. In the third stage (200ps-500ps), the Cu and C diffusion coefficients remain close values as below ~0.02 $Å^2$/ps, indicating atomic diffusion has been constrained under such a low temperature.

The time-dependent radial distribution functions (RDF) shown in Fig. 2a have recorded the evolution of C-C and Cu-Cu atomic distance during phase separation, with the first nearest neighbor of 1.38Å and 2.49Å at equilibrium state, respectively. Fig. 2b gives the statistical distribution of Cu-Cu coordination with equilibrium coordination number ranging from ~10 to ~20 by using the Voronoi analysis of OVITO. Notably, we use the order parameter of q4 and q6 introduced by Steinhardt *et al*. to characterize the local bond-orientational order of Cu and C atoms [27]. As shown in Fig. 2c, the majority of bond order towards Cu exhibits a higher value of q6 than C.

As one of the critical factors that may influence atomic distribution during phase separation, the pressure during atomic deposition should be quantitatively studied. Fig. 3 a-c depict the time-dependent snapshots of Cu crystalline morphology with simulated pressure ranging from 1 kPa to 1 GPa. Assuming the deposited Cu and C atoms are randomly distributed on the substrates under 1000 K at the initial stage (time=0). As the temperature decreases, the diffusive atoms will spontaneously form clusters (time=100ps). The emerging spherical Cu crystals are distinctively separated from the C atoms once the quenching time exceeds 500ps. This morphology is consistent with the as-synthesized nanocrystalline-amorphous structures that observed by HRTEM. Notably, if increasing the pressure from 1 kPa to 1 MPa, the growth of spherical Cu crystalline is dramatically promoted, while the crystalline will be evolved into the spot shape once the pressure rises to 1 GPa. The initial Cu content is the other critical parameter during phase separation. As presented in Fig. 3 d-f, the Cu crystalline's average size will rise with increasing Cu concentration, theoretically explaining the growth of Cu nanocrystalline with decreasing $C_2H_2$ gas content.

More specific results about statistical grain size by using the Polyhedral analysis of OVITO under different pressure and Cu concentration is given in Fig. 4 a-d. By defining the initial concentration with Cu:C=1:1, the trend of Cu crystal ratio strongly depends on pressure with the largest value at MPa (Fig. 4a). If keep constant pressure at 1 GPa, the Cu crystal ratio will approximately exhibit linear growth with increasing Cu concentration (Fig. 4b). Similar conclusion can be obtained by converting crystal ratio with summation crystal size as presented in Fig. 4c. Fig. 4d also shows an increasing trend of average crystal size with Cu concentration ratio rising from 0.6 to 0.8. These results highlight both the pressure and initial Cu concentration play a significant role in the final crystalline morphology and size during phase separation.

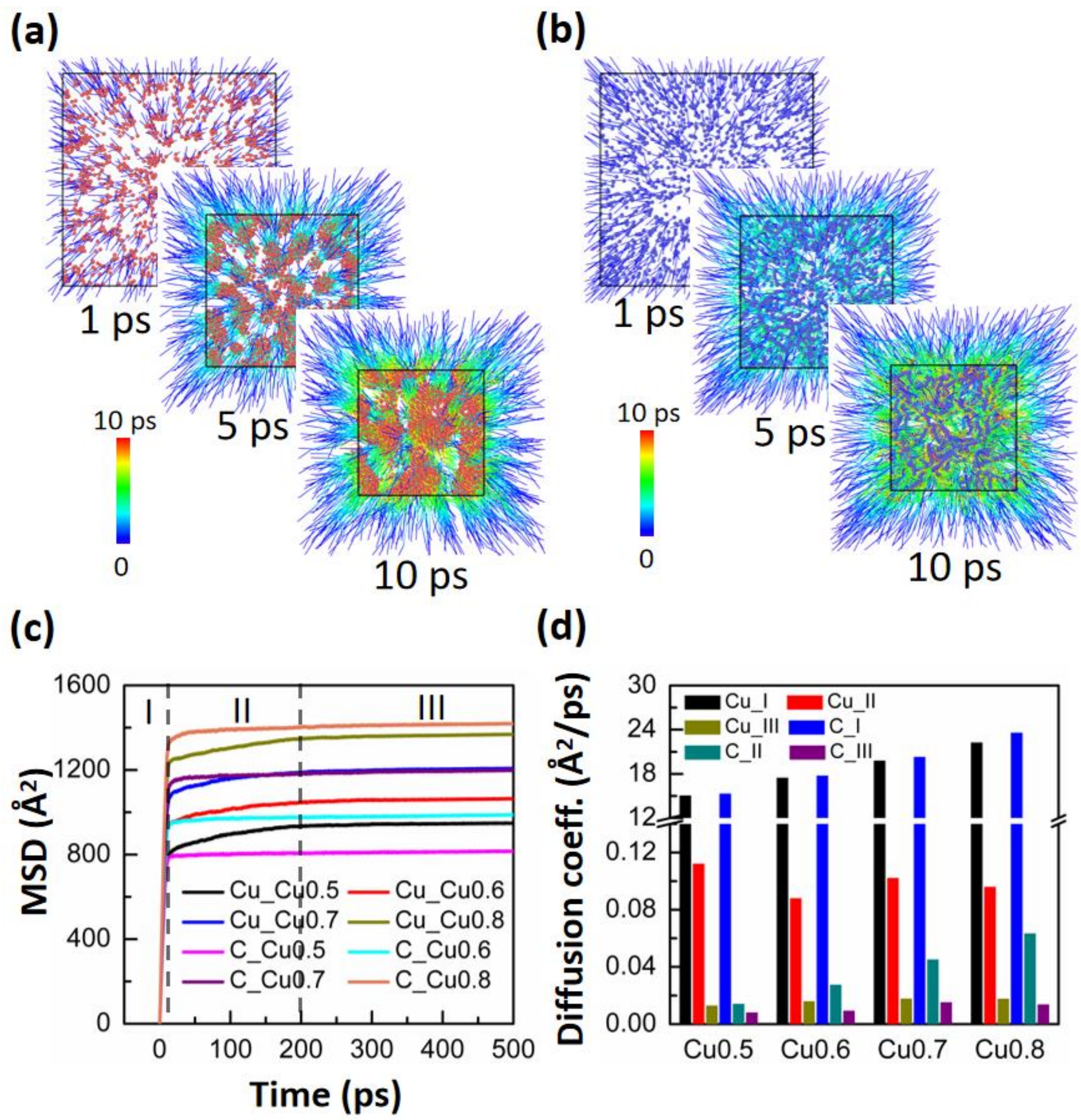

**Figure 1**. Snapshots of diffusion trajectory line of (a) Cu and (b) C atoms with initial Cu:C=1:1 quenching from 1000K to 300K under 1GPa; the MSD (c) and corresponding diffusion coefficient (d) of Cu and C.

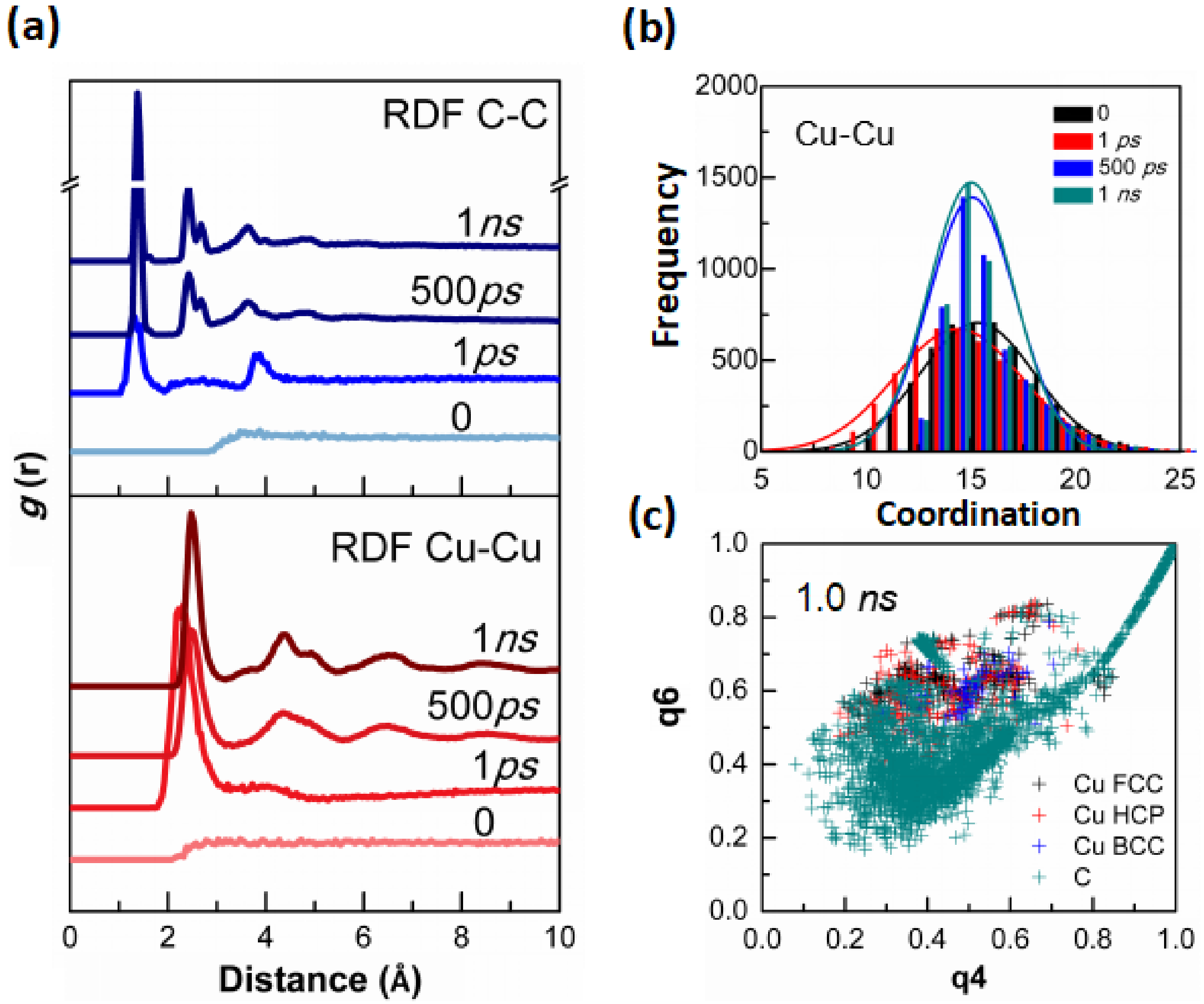

**Figure 2**. (a) The RDF of C-C and Cu-Cu with initial Cu:C=1:1 quenching from 1000K to 300K under 1GPa; (b) Cu-Cu coordination; (c) the order parameter q4 and q6 of Cu and C.

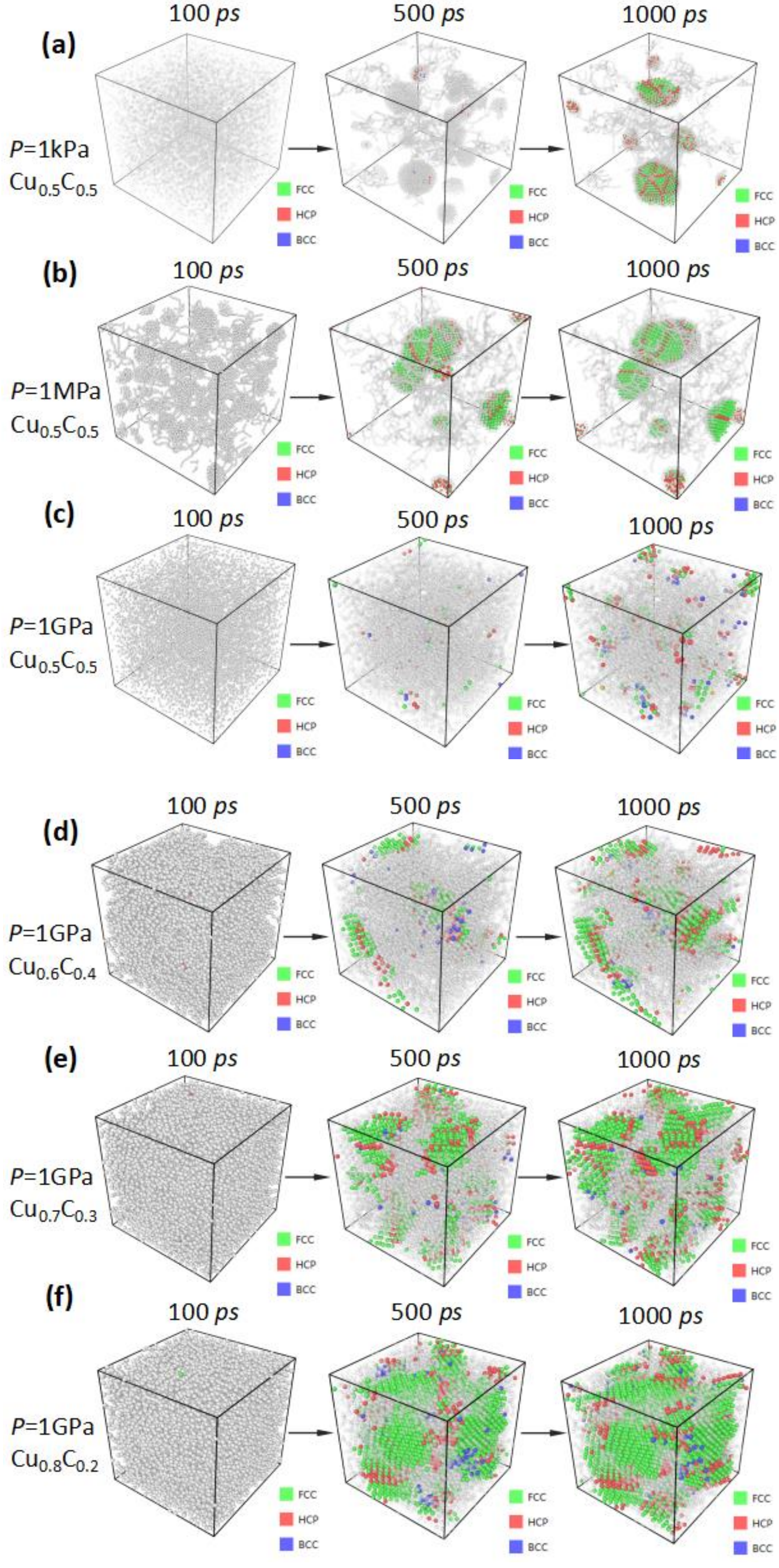

**Figure 3**. (a)-(c) Snapshots of Cu crystalline with initial Cu:C=1:1 quenching from 1000K to 300K under different pressure; (d)-(f) Snapshots of Cu crystalline with different initial Cu content quenching from 1000K to 300K under 1GPa.

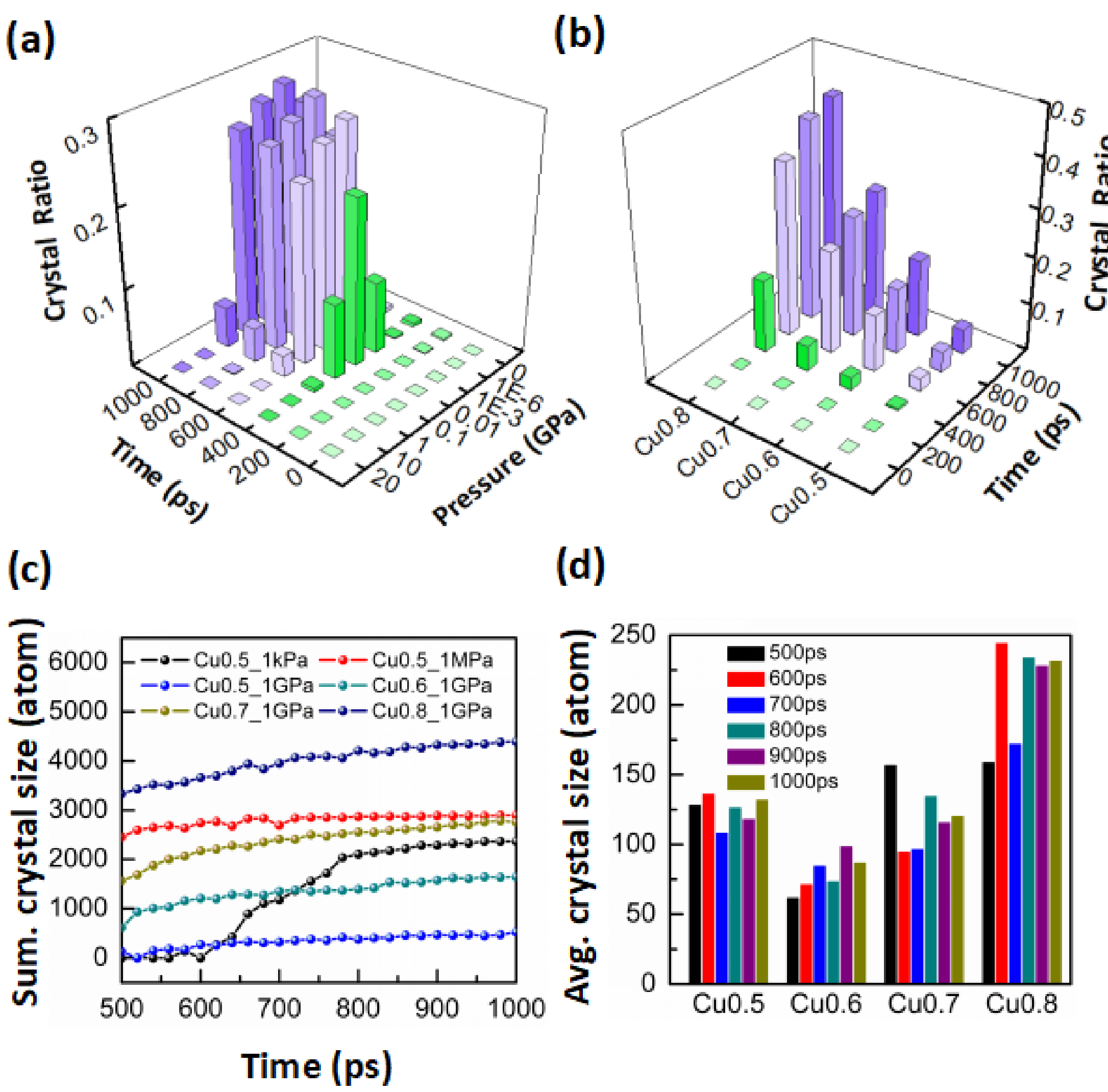

**Figure 4**. Cu crystal ratio with different initial (a) pressure and (b) Cu content quenching from 1000K to 300K; evolution of Cu grain size with different initial (c) pressure and (d) Cu content.

## IV. CONCLUSIONS

In this work, a binary Cu-C phase separation was investigated by using classical MD simulation. In order to unveil the key factors that can fundamentally determine the final phase structures, we used the NPT ensemble under various concentration ratio and pressure. The simulation theoretically reproduces the Cu-C three possible diffusion stages and final phase separation, and reveals that initial Cu concentration and pressure play significant roles in the evolution of Cu crystal's morphology and size. The pressure at MPa level will contribute to the largest Cu crystal ratio. On the other hand, a relatively high initial Cu concentration facilitates rapid crystal growth during phase separation. We hope this finding may provide a profound understanding in nanocrystalline-amorphous phase separation and present practical methodology for depositing functional coatings with designed microstructures and properties.

## ACKNOWLEDGMENT

This work is partly supported by the National Natural Science Foundation of China (No.12175019), the National Natural Science Foundation Joint Fund Key Project (U1865206), the Technological Innovation and Application Development (general Project) (cstc2020jscx-msxm1818) in Chongqing, and China Scholarship Council (CSC)

## REFERENCES

(1) G. Radnóczi, E. Bokányi, Z. Erdélyi, and F. Misják, "Size dependent spinodal decomposition in Cu-Ag nanoparticles," Acta Mater. **123**, 82-89 (2017).

(2) M. Powers, B. Derby, A. Shaw, E. Raeker, and A. Misra, "Microstructural characterization of phase-separated co-deposited Cu-Ta immiscible alloy thin films," J. Mater. Res. **35**, 1531-1542 (2020).

(3) B. Derby, Y. Cui, J. K. Baldwin, and A. Misra, "Effects of substrate temperature and deposition rate on the phase separated morphology of co-sputtered, Cu-Mo thin films," Thin Solid Films **647**, 50-56 (2018).

(4) G. Abrasonis, G. J. Kovάcs, M. D. Tucker, R. Heller, M. Krause, M. C. Guenette, F. Munnik, J. Lehmann, A. Tadich, B. C. C. Cowie, L. Thomsen, M. M. M. Bilek, and W. Möller, "Sculpting nanoscale precipitation patterns in nanocomposite thin films via hyperthermal ion deposition," Appl. Phys. Lett. **97**, 163108 (2010).

(5) G. Abrasonis, Gy. J. Kovács, L. Ryves, M. Krause, A. Mücklich, F. Munnik, T. W. H. Oates, M. M. M. Bilek, and W. Möller, "Phase separation in carbon-nickel films during hyperthermal ion deposition," J. Appl. Phys. **105**, 083518 (2009).

(6) P. Eh. Hovsepian, Y. N. Kok, A. P. Ehiasarian, R. Haasch, J.-G. Wen, and I. Petrov, "Phase separation and formation of the self-organised layered nanostructure in C/Cr coatings in conditions of high ion irradiation," Surf. Coat. Technol. **200**, 1572-1579 (2005).

(7) W. Wesch, and E. Wendler, in *Ion Beam Modification of Solids: Ion-Solid Interaction and Radiation Damage* (Springer, 2016), pp. 64-79.

(8) S. Veprek, "The search for novel, superhard materials," J. Vac. Sci. Technol. **17**, 2401-2420 (1999).

(9) S. Veprek, "New development in superhard coatings: the superhard nanocrystalline-amorphous composites," Thin Solid Films **317**, 449-454 (1998).

(10) J. W. Cahn, and J. E. Hilliard, "Free Energy of a Nonuniform System. I. Interfacial Free Energy," J. Chem. Phys. **28**, 258-267 (1958).

(11) J. W. Cahn, "Phase Separation by Spinodal Decomposition in Isotropic Systems," J. Chem. Phys. **42**, 93 (1965).

(12) T. Barkar, L. Höglund, J. Odqvist, and J. Ågren, Comp. "Effect of concentration dependent gradient energy coefficient on spinodal decomposition in the Fe-Cr system," Mater. Sci. **143**, 446-453 (2018).

(13) B. König, O. J. J. Ronsin, and J. Harting, "Two-dimensional Cahn-Hilliard simulations for coarsening kinetics of spinodal decomposition in binary mixtures," Phys. Chem. Chem. Phys. **23**, 24823-24833 (2021).

(14) P. K. Inguva, P. J. Walker, H. W. Yew, K. Z. Zhu, A. J. Haslam, and O. K. Matar, "Continuum-scale modelling of polymer blends using the Cahn-Hilliard equation: transport and thermodynamics," Soft Matter **17**, 5645-5665 (2021).

(15) Y. Lu, C. P. Wang, Y. P. Gao, R. P. Shi, X. J. Liu, and Y. Z. Wang, "Microstructure map for self-organized phase separation during film deposition," Phys. Rev. Lett. **109**, 086101 (2012).

(16) K. Ankit, B. Derby, R. Raghavan, A. Misra, and M. J. Demkowicz, “3-D phase-field simulations of self-organized composite morphologies in physical vapor deposited phase-separating binary alloys,” J. Appl. Phys. **126**, 075306 (2019).
(17) J. A. Stewart, and R. Dingreville, “Microstructure morphology and concentration modulation of nanocomposite thin-films during simulated physical vapor deposition,” Acta Mater. **188**,181-191 (2020).
(18) G. Kairaitis, and A. Galdikas, “Phase separation during thin film deposition,” Comp. Mater. Sci. **91**, 68-74 (2014).
(19) P. F. Hu, and B. L. Jiang, “Study on tribological property of CrCN coating based on magnetron sputtering plating technique,” Vacuum **85**, 994-998 (2011).
(20) J. J. Guan, H. Q. Wang, L. Z. Qin, B. Liao, H. Liang, and B. Li, “Phase transitions of doped carbon in CrCN coatings with modified mechanical and tribological properties via filtered cathodic vacuum arc deposition,” Nucl. Instrum. Methods Phys. Res. **397**, 86-91 (2017).
(21) P. B. Barna, M. Adamik, J. Lábár, L. Kövér, J. Tóth, A. Dévényi, and R. Manaila, “Formation of polycrystalline and microcrystalline composite thin films by codeposition and surface chemical reaction,” Surf. Coat. Technol. **125**, 147-150 (2000).
(22) A. P. Thompson, H. M. Aktulga, R. Berger, D. S. Bolintineanu, W. M. Brown, P. S. Crozier, P. J. in 't Veld, A. Kohlmeyer, S. G. Moore, T. D. Nguyen, R. Shan, M. J. Stevens, J. Tranchida, C. Trott, and S. J. Plimpton, “LAMMPS-a flexible simulation tool for particle-based materials modeling at the atomic, meso, and continuum scales,” Comp Phys Comm. **271**, 10817 (2022).
(23) S. Plimpton, “Fast Parallel Algorithms for Short-Range Molecular Dynamics,” J Comp Phys, **117**, 1-19 (1995).
(24) X. W. Zhou, H. N. G. Wadley, R. A. Johnson, D. J. Larson, N. Tabat, A. Cerezo, A. K. Petford-Long, G. D. W. Smith, P. H. Clifton, R. L. Martens, and T. F. Kelly, “Atomic scale structure of sputtered metal multilayers,” Acta Mater, **49**, 4005 (2001).
(25) S. J. Stuart, A. B. Tutein, and J. A. Harrison, “A reactive potential for hydrocarbons with intermolecular interactions,” J. Chem. Phys. **112**, 6472-6486 (2000).
(26) A. Stukowski, “Visualization and analysis of atomistic simulation data with OVITO-the Open Visualization Tool,” Modelling Simul. Mater. Sci. Eng. **18**, 015012 (2010).
(27) P. J. Steinhardt, D. R. Nelson, and M. Ronchetti, “Bond-orientational order in liquids and glasses,” Phys. Rev. B **28**, 784 (1983).